\documentstyle[epsf,twocolumn,prb,aps]{revtex}

\begin{document}
\draft

\title{Normalization factors 
for magnetic relaxation of small particle systems 
in non-zero magnetic field.}

\author{$^{1}$Ll. Balcells, $^{2,}$\cite{Author} O. Iglesias, 
and $^{2}$A. Labarta} 

\address{
$^{1}$Institut de Ciencia de Materials de Barcelona-CSIC, 
Campus de la UAB, 08193 Bellaterra \\
$^{2}$Departament de F\'{\i}sica Fonamental, Facultat de F\'{\i}sica, 
Universitat de Barcelona, Diagonal 647, 08028 Barcelona, Spain
}

\date{Last version \today}
\maketitle

\newcommand{\svar}{T \ln(t/\tau_0)} 

\begin{abstract}
We critically discuss relaxation experiments in magnetic systems 
that can be characterized in terms of an energy barrier distribution,  
showing that proper normalization of the relaxation data is needed 
whenever curves corresponding to different temperatures are to be 
compared. We show how these normalization factors can be 
obtained from experimental data by using the $\svar$ scaling method 
without making any 
assumptions about the nature of the energy barrier distribution. 
The validity of the procedure is tested using a ferrofluid of 
$\rm Fe_3 \rm O_4$ particles.   
\end{abstract}


\pacs{PACS numbers: 75.50.Tt, 75.60.Lr.}
\section{Introduction}
\label{introduction}

The study of the relaxation of magnetic systems provides a way 
to obtain information about different properties 
that cannot be so easily achieved by other methods. 
Most of the works in this field are based on the logarithmic 
or critical volume approximation \cite{Bean}. To characterize the time
dependence of the magnetization they analyze the relaxation rate, 
also called magnetic viscosity, $S$, as a function of the external parameters. 
When plotted as a function of the magnetic field it is possible 
to study the variation of the energy barriers with the field \cite{Barrfield}, 
the interaction effects among the magnetic entities \cite{Rome,Montse},  
and the switching field distribution \cite{Spratt,Ogrady} 
among other magnetic properties.
When plotted as a function of the temperature for a given field it
gives information about the reversal mechanisms in films and small 
particle systems \cite{Reversal} and it has been proposed as a way 
to test the possibility 
of observing macroscopic quantum tunneling (MQT) effects at 
low enough temperatures \cite{QTM}. It is obvious that 
in this case the initial and final states of the relaxing  
magnetization are in general temperature dependent. Therefore,  
this dependence must be corrected in $S$ data in order to 
compare values obtained at different temperatures. 
If this correction is not taken into account this trivial contribution 
to the thermal dependence of $S$ can obliterate the real $S(T)$ behaviour  
arising from the relevant physical mechanism and it can even give rise 
to misleading interpretations. 

Moreover, in some cases, $\svar$ scaling has been used \cite{QTM2,Vincent} to  
confirm the existence of MQT by adducing that the fail of scaling of 
low temperature relaxation curves could be the signature of non-thermal 
mechanisms. 
As we will show later this lack of scaling could be only a consequence 
of a non-proper normalization of the data.

In most of particle systems in the 
blocked regime and due to the long-time decay towards the equilibrium 
state, it is very difficult to obtain a direct experimental
determination of the equilibrium magnetization when a magnetic field 
is applied (note that, in general, the field-cooled state is not a 
true equilibrium state). 
It is the purpose of this article to give 
a detailed account of the normalization procedure 
of the relaxation curves based on the so-called $\svar$ scaling method 
\cite{Prejean,PRB93} from which an indirect determination of the thermal 
dependence of the equilibrium magnetization (final state) can be obtained.  
We also discuss the consequences of this normalization 
procedure on the analysis of the $S(T)$ 
behaviour. The proposed method is illustrated by using experimental data 
from a ferrofluid composed of $\rm Fe_3 \rm O_4$ small particles.    

\section{Normalization factors for relaxation curves}
\label{relaxation}

In the study of time dependent processes in small particles systems
two kind of experiments (which will be called type A and B) 
can be distinguished according to what is the 
final equilibrium state of the system.  
In Type A experiments the system evolves towards 
a demagnetized state in zero applied field after a previous cooling 
in the presence of a field $H$ (FC process) and the variation of the
thermoremanent magnetization (TRM) is measured.
If $t$ is the time 
elapsed after the field was reduced to zero then, in the critical
volume approximation \cite{Ogrady}:
\begin{eqnarray}
M_{TRM}(T,H,t)=
\int _{E_c(T,t)} ^ \infty dE {M}_{FC}(E)f(E)\nonumber\\ 
\simeq {M}_{FC}(T,H) \int _{E_c(T,t)} ^\infty  dE f(E).
\label{m4}
\end{eqnarray}
where $E_c$ is the critical energy barrier $E_c(T,t)=T\ln(t/\tau_0)$ 
which indicates the onset of superparamagnetic (SP) behaviour. 

In Type B experiments a zero-field cooled (ZFC) sample 
increases its magnetization 
in a magnetic field $H$ and the variation of the isothermal 
remanent magnetization (IRM) is measured. 
In the critical volume approximation the time dependence of the 
magnetization is given in this case by \cite{Ogrady}:
\begin{eqnarray}
M_{IRM}(T,H,t)=
\int _0 ^ {E_c(T,t,H)} dE {M}_{eq}(E) f(E)\nonumber\\ 
\simeq {M}_{eq}(T,H) \int _0 ^ {E_c(T,t,H)}dE f(E).
\label{m5}
\end{eqnarray}

In order to compare relaxation curves 
measured at different temperatures it is necessary to 
remove the thermal dependence of the initial and final states of 
the magnetization. It is clear from Eq. \ref {m4} and \ref {m5} 
that, due to the fact that in both 
expressions the integrals are bounded between 0 and 1, 
this can be simply achieved by dividing magnetization data 
by a normalization factor which in Type A experiments is ${M}_{FC}(T,H)$
and in Type B is ${M}_{eq}(T,H)$.

In Type A experiments, the
normalization factor, $M_{FC}(T)$, in (\ref{m4}) comes from the 
contribution of the blocked particles to the initial magnetization 
or, in other words, from the irreversible component of the FC magnetization 
at the temperature $T$. In many small particle systems at low $T$ 
this quantity can be considered as a constant in 
the range of temperatures usually studied \cite{Rome}. In fact, when one 
represents the relaxation data as a function of the scaling variable $\svar$ 
all the curves recorded at different temperature usually superimpose onto a 
unique master curve without any normalization factor \cite{PRB93}.

In contrast, in Type B experiments  
the magnetic field is not zero 
and SP particles have a temperature dependent contribution 
to the magnetization while blocked particles are randomly 
oriented giving no net contribution to the magnetization in 
the field direction. 
Now the reversible component of the ZFC magnetization gives the 
main contribution to the normalization factor, ${M}_{eq}(T)$. In many cases, 
at low fields, ${M}_{eq}(T)$ follows a 
Curie-Weiss law and cannot be considered as a constant. The need of 
this temperature dependent normalization factors is clearly manifested   
as a vertical shift of the 
curves when data are represented in a $\svar$ scaling plot.

\section{Experimental results}
\label{exp}

The studied sample is a ferrofluid composed of ${\rm Fe}_3\rm{O}_4$ 
small particles with volume packing fraction $\epsilon=0.01$ 
which was obtained from chemical deposition of 
$\rm{Fe}^{+2}$ and $\rm{Fe}^{+3}$ sulfides 
and dispersed in an hydrocarbon oil. 
The sample analyzed by transmission electron microscopy (TEM) was prepared 
by wetting a carbon film mounted 
on an Au grid with the ferrofluid and subsequently drying it with air.

In Fig. \ref{microgr} an electron micrograph of the
magnetic particles is shown. Taking a sampling of 200 particles and
considering them spherical, the particle size distribution 
have been fitted to a logarithmic-normal function with $\sigma
= 0.24 $ and mean particle diameter of about 4 nm (see inset 
of Fig. \ref{microgr}).


\subsection{Magnetic characterization}

The magnetic study was performed with a commercial SQUID magnetometer. 
The highest applied field was 55 kOe and the lowest
temperature was 1.8 K.

Figure \ref{zfcfc} shows the thermal dependence of the magnetization 
in an applied magnetic field of 10 Oe following a ZFC-FC process.
The magnetic behaviour displayed is typical of SP particles. 
Above $T_{irr}=50$ K the system
is in the SP region, in which the magnetization curve is
reversible. The ZFC curve developes a maximum at $T_B$ which is 
about 14 K. 

To obtain the mean magnetic moment of the particles  
as a function of the temperature, $\bar{\mu}(T)$, we have fitted
the magnetization curves deep in the SP region (within the range 80 
to 200 $K$)
to a distribution of Langevin functions following the procedure  
described in Ref. \onlinecite{Linder}.
By extrapolating these values to $T=0$ K with a $T^{3/2}$ law
(as expected according to a spin-wave theory for a 
ferrimagnetic material \cite{Ashcr}), 
a mean magnetic moment per particle of $\bar{\mu}=
(1320\pm 20) \mu_B$ is obtained. 
Using $\bar{\mu}$ and the bulk magnetic moment of the
magnetite the mean magnetic size of the particles has been found
to be 3.5 nm in diameter. 

$T_B$ has been measured as a function of the 
field. By extrapolating these values to $T_B=0$
we have estimated the mean anisotropy field 
(the field at which the mean energy barrier disappears)  
to be $H_A \simeq 5000$ Oe. 
From $H_A$ the mean anisotropy constant of
the particles have been estimated as 
$K=\bar \mu H_A/2 \bar V= 1.3 \times 10^5 {\rm J/m}^3$, where $\bar V$ 
is the mean particle volume. 
This value is close to others found in the literature for similar
systems \cite{Jonsson,Luo,Johansson}.

For the subsequent analysis of the relaxation curves it is
also convenient to study the variation of the initial susceptibility 
with the temperature. For a system of interacting 
magnetic particles in the SP
regime this quantity can be written as \cite{Elhilo}:
\begin{equation}
\chi \sim {\bar \mu^2\over 3k_B(T - T_0)}
\label{chi}
\end{equation}
where $T_0$ is due to the existence of interparticle interactions 
and/or the effect of $f(E)$ (see Ref. \cite{Elhilo}). 
The inset in Fig. \ref{zfcfc} shows the inverse susceptibility as a
function of the temperature as obtained from the ZFC-FC curve, where the  
thermal variation of the mean moment of the particles has been corrected 
by using the $T^{3/2}$ law obtained before. Above
 $T_{irr}$, $1/\chi$ shows a linear dependence on $T$. 
Fitting $1/\chi$ to (\ref{chi})   
we have estimated $T_{0}=-11 \pm 3$ K. This value of $T_{0}$ may be 
mainly attributed to the existence of demagnetizing interparticle 
interactions taking into account that the volume 
distribution is not very broad.


\subsection{Normalization factors and $\bf {T\ln (t\tau_0)}$ scaling}
\label{norm}

The magnetic relaxation curves at different temperatures were 
recorded with a SQUID magnetometer following the
procedure described in Ref. \onlinecite{Labarta}.
The measurements were 
started 5 s after applying the field and were performed during aprox. 
1000 s at the 
lowest temperature and aprox. 10000 s at the highest temperature. 
The relaxation curves after ZFC the system were measured 
in the presence of a magnetic field of 10 Oe (type B experiment) 
while in those measured 
in zero field the system was previously FC in 10 Oe (type A experiment). 

In the following we will describe an experimental procedure, 
based on the $\svar$ scaling method \cite{PRB93}, to obtain 
normalization factors, $M_{eq}(T)$, for relaxation data 
recorded in the presence of a field since they cannot be 
directly measured due to the long-time decay of the magnetization.

Firstly, the attempt time $\tau_0=3\times 10^{-11}$ s 
has been evaluated by scaling the relaxation curves in 
zero field after FC the system at several temperatures following 
the method previously described in Ref. \onlinecite {PRB93}.  
For this purpose we have considered that the initial magnetization 
${M}_{FC}(T,H)$ is constant in the temperature range of the experiment 
so no normalization factors are needed to scale the curves. 
We will assume that the field variation of $\tau_0$ 
is smaller than the error in its determination at low fields, 
according to Brown's theory \cite{Brown}.

In the next step, relaxation data recorded in the presence of a field 
have been plotted as a function of 
the scaling variable using the 
value of $\tau_0$ previously deduced (see Fig. \ref{relax}). 
After this transformation the relaxation curves at different 
temperatures are separated 
along the vertical axis by temperature dependent shifts.
Taking into account that the applied field is much lower than $H_A$ 
and therefore the energy barriers have not been significatively affected, 
this lack of scaling is a clear demonstration that magnetization 
data must be normalized to achieve an equivalent scaling to that 
obtained in zero field.    
As has been discussed in Sec. \ref{relaxation} the normalization 
factors are proportional to $M_{eq}(T,H)$. If now we assume that 
$M_{eq}(T,H)$ are given by $M_{FC}(T,10$ Oe), as suggested by some 
authors \cite{Vincent}, no scaling is achieved because FC magnetization
does not correspond to the true equilibrium state.
Note that $M_{eq}(T,H)$ can not be calculated 
without making any a priori hypothesis about the form of $f(E)$ and the 
magnetic microstructure of the system. 

The normalization factors can be found by referring 
the different curves to the lowest temperature one. 
Once this process has been performed, the relaxation curves  
collapse onto a unique master curve that is shown in Fig. \ref{mastcurve}.  
The values of the normalization factors  
follow a Curie-Weiss law of the form (\ref{chi}) 
with $T_{0}=-15 \pm 2$ K (see inset of Fig. \ref{factors}). 
The extrapolation of this law superimposes with the 
susceptibility corresponding to the reversible (SP) region obtained from 
ZFC-FC measurements (see Fig. \ref{factors}), reflecting the fact 
that for long enough observation times all the particles 
have become SP and demonstrating that the normalization factors are 
proportional to $M_{eq}(T,H)$.


\subsection{Magnetic viscosity and energy barrier distribution}

The classical magnetic viscosity commonly defined as 
$S(t,T)= {\partial M(t) / \partial (\ln t)}$, 
cannot be directly compared at different temperatures because:  
1) usually magnetization is not normalized (initial and final states 
of the relaxation process change as the temperature varies)
and 2) if relaxation data have been recorded in a fixed time window 
the energy barriers which are relaxing at different temperatures are 
not the same.
Both problems can be circumvented by defining magnetic viscosity as 
$\bar{S} (t,T)= {\partial \bar{M} (t) / \partial (\svar)}$ ($\bar{M}(t)$ 
is the normalized magnetization used in the scaling procedure). 
By performing the $\svar$ derivative $\bar S$ measures the relaxation rate  
of the magnetization due only to the energy barriers around $\svar$. 
On the other hand, as previously noted in Ref.\onlinecite{ZPB96}, 
$\bar{S}$ is a magnitude proportional to the energy barrier 
distribution and therefore it has a direct physical meaning.
 
These two magnitudes are simply related by
\begin{equation}
\bar{S} ={S \over M_{FC} T}
\label{s1}
\end{equation}
in the case of type A relaxation experiments, and by
\begin{equation}
\bar{S} ={S \over M_{eq} T}
\label{s2}
\end{equation}
for type B experiments. In the first case, and for systems with 
a certain degree of interaction between particles $M_{FC}$ is usually 
almost temperature independent 
and both magnitudes differ by a $T^{-1}$ factor. Therefore 
conclusions from the thermal variation of $S$ obtained from type A 
experiments must be carefully derived. Note 
in particular that if $S$ happens to be temperature independent in a 
certain range, a result 
which could be interpreted as a proof of the existence of 
quantum relaxation phenomena \cite{Qrelax}, this would be a consequence of 
an energy barrier distribution proportional to $1/E$ in this range, as 
the $\bar S \sim T^{-1}$ thermal dependence reveals. It is worth        
noticing that if there is not a certain degree of freezing due to magnetic
interactions, $M_{FC}$ can not be considered as a constant and its 
thermal variation must be corrected in $S$.
However, in type B experiments, only if $M_{eq}(T)$ is 
inversely proportional to the temperature (Curie law), 
as is the case for a sample with no or very small interparticle 
interactions and a narrow distribution of energy barriers \cite{Elhilo}, 
both magnitudes nearly coincide (note that this is not the case 
of the sample studied in this paper) because the thermal variation of 
$M_{eq}$ cancels the factor T in Eq. \ref{s2}.

For the sample studied in this paper $\bar S(\svar)$ has been 
obtained by performing the numerical derivative of the master curve 
of Fig. \ref{mastcurve} and has been compared to the viscosity $S(T)$ 
as obtained from the logarithmic derivative of the relaxation data 
at each temperature. The results are shown in Fig. \ref{effdistr} together 
with the energy distribution obtained from the electron micrography 
by expressing the volume distribution in energy units with the help of the  
value of $K$ previously derived. The coincidence between $\bar S$ and $f(E)$ 
shows the consistence of the normalization used in the scaling 
procedure for type B experiments. 
On the other hand, it is important to note that $S$ does not coincide 
with $\bar S$ because for this sample $M_{eq}$ is not simply 
proportional to $T^{-1}$. Only the overall shape of the energy barrier 
distribution obtained from TEM is reproduced by $S$, but shifted to higher 
energies. Note also that the extrapolation of the quasi-linear 
low temperature regime of $S$ intercepts the temperature axis at a 
non-zero value as has been reported in other systems \cite{}. 
In our case, this result is only a consequence of the 
lack of normalization and has no physical meaning.  

\section{conclusions}
\label{conclusions}

We have stressed the importance of proper normalization whenever 
relaxation curves measured at different temperatures must be compared. 
In the case of experiments performed in zero field, 
care must be taken in systems for which $M_{FC}(T)$ cannot be 
considered as a constant. When this is the case, non-normalization 
could give place to a spurious thermal dependence. 

In the case of relaxation experiments performed in an  
applied magnetic field, there exists a certain controversy in the 
literature about the nature of the normalization : either no normalization 
factors are used at all \cite{QTM,Perera,Ibrahim} 
or the FC magnetization value, corresponding 
to the field at which the experiment is performed \cite{Vincent}, is used. 

In systems for which the $M_{eq}(T)$ follows a Curie law (non-interacting 
particles, neglictible $T_{0}$) the first option happens to be correct 
by chance as can be easily seen in (\ref{s2}). In this kind of systems
the second option is particularly wrong when applied to low temperature data 
because the FC magnetization is slightly temperature dependent while 
the SP magnetization, which is the true equilibrium state at long times, 
follows a $T^{-1}$ behaviour.

In fact, when $\svar$ scaling is used to evidence quantum relaxation 
mechanisms through a lack of scaling of relaxation data (recorded 
in the presence of a field), no conclusions should be extracted without 
previously having tried to normalize data following the process described 
in Sec. \ref{norm}. The sample studied in this paper is an example 
where a clear fail of scaling of the non-normalized data does not 
indicate any non-thermal process ($\bar S$ agrees with the energy 
distribution deduced from $f(V)$, see Fig. \ref{effdistr}). 
This does not mean that it is always possible to find scaling factors 
for data corresponding to low $T$ if MQT occurs. If this is the case,
even a multiplicative factor is not enough to superimpose relaxation
curves obtained at different $T$ ($T$ is then an irrelevant parameter 
that would not have to be included in the scaling variable). 

In conclusion, we have shown that the $\svar$ scaling method 
provides a useful tool to 
obtain the normalization factors and the energy barrier distribution 
in both kinds of experiments and even in systems that cannot be 
considered as an assembly of independent small particles
(i.e. multilayered systems, cluster glasses, amorphous alloys, etc.)   
without making any assumptions about the nature of the magnetic 
microstructure.


\acknowledgements
We are indebted to professor S.W. Charles for providing the sample studied 
in this paper.
Finantial support from both the Spanish CICYT through MAT94-1024-CO2-02
and the Catalan CIRIT through GRQ1012 is acknowledged.



\begin{figure}
\caption{Electron micrography of the sample obtained by TEM. 
The inset shows the distribution of particle diameters 
obtained from a sampling of 200 particles. 
The solid line is a logarithmic-normal function with 
$\sigma = 0.24$ and mean particle diameter of 
about 4 nm.
}
\label{microgr}
\end{figure}

\begin{figure}
\caption{Temperature dependence of the magnetization of the sample for a 
ZFC (lower curve)- FC (upper curve) process in a magnetic field of 10 Oe.
Inset: reciprocal of the ZFC-FC susceptibility corrected 
from the thermal variation of the mean particle moment.  
}
\label{zfcfc}
\end{figure}

\begin{figure}
\caption{Relaxation data recorded from 1.8 to 15 K in the 
presence of a magnetic field of 10 Oe after ZFC the sample
as a function of the $\svar$ scaling variable. Open and solid 
simbols correspond alternatively to the 
temperatures indicated in the Figure. 
Inset: detail of the lowest
temperature region. 
}
\label{relax}
\end{figure}

\begin{figure}
\caption{Scaling plot for the relaxation measurements shown in 
Fig. \ref{relax}. 
Open and solid simbols correspond alternatively to the 
temperatures indicated in the figure. 
Inset: detail of the lowest temperature region.
}
\label{mastcurve}
\end{figure}

\begin{figure}
\caption{Reversible region of the reciprocal susceptibility (open squares) 
and thermal dependence of the inverse of the normalization factors 
(solid circles) necessary 
to join the relaxation data of Fig. \ref{relax} onto a unique 
master curve. Solid line is a linear regression of both data. 
Inset: inverse of the normalization factors as a 
function of the temperature. Normalization factors have been reduced to the same 
units of the susceptibility multiplying them by an arbitrary quantity.
}
\label{factors}
\end{figure}

\begin{figure}
\caption{Effective distribution of energy barriers as obtained from the
numerical derivative of the master relaxation curve (solid line). 
Magnetic viscosity $S(T)$ as obtained from the logarithmic 
time derivative of the relaxation data at the temperatures 
indicated in Fig. \ref{mastcurve} (solid squares).
The energy distribution $f(E)$ obtained from $f(V)$  
is also shown for comparison (open circles).
}
\label{effdistr}
\end{figure}
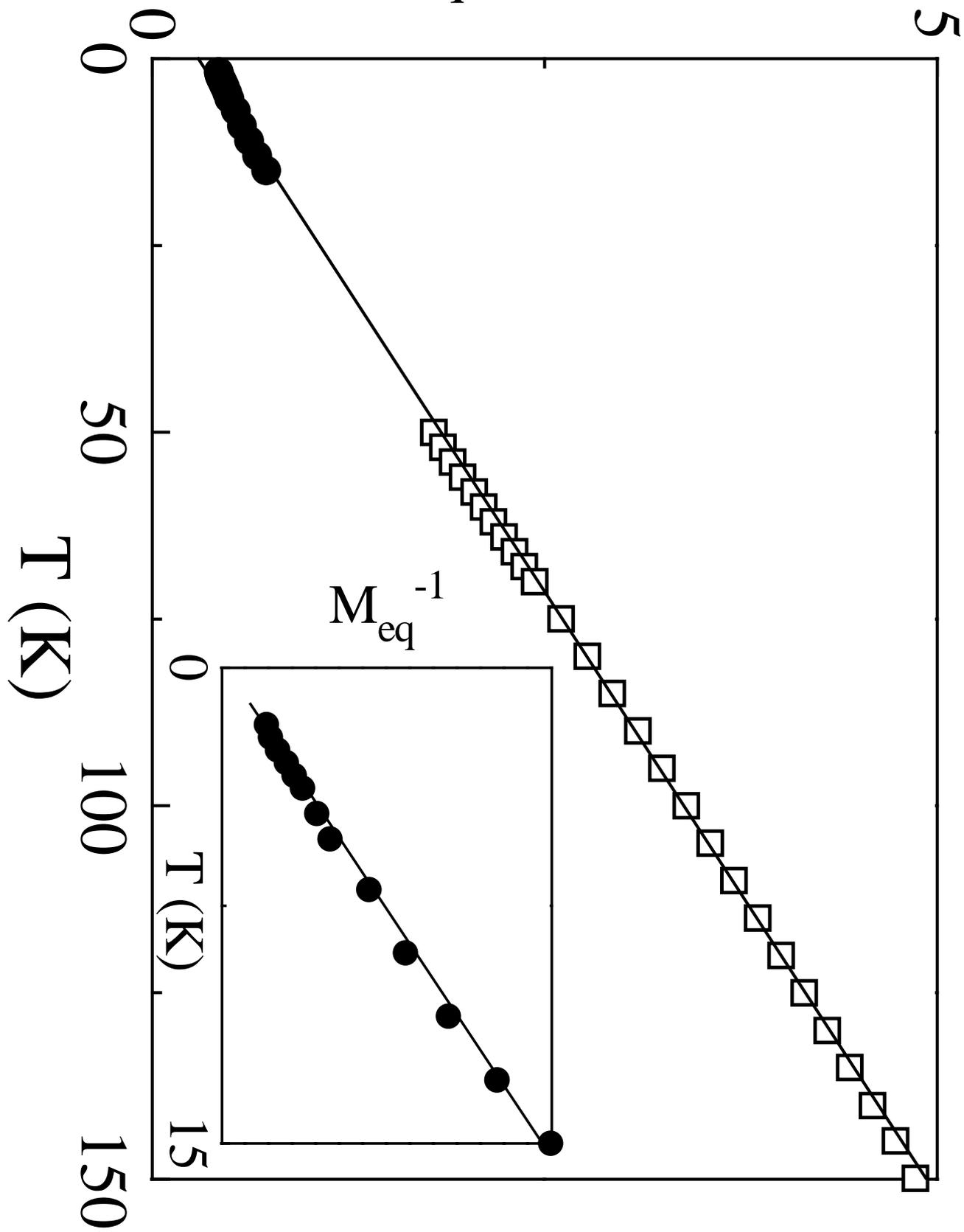
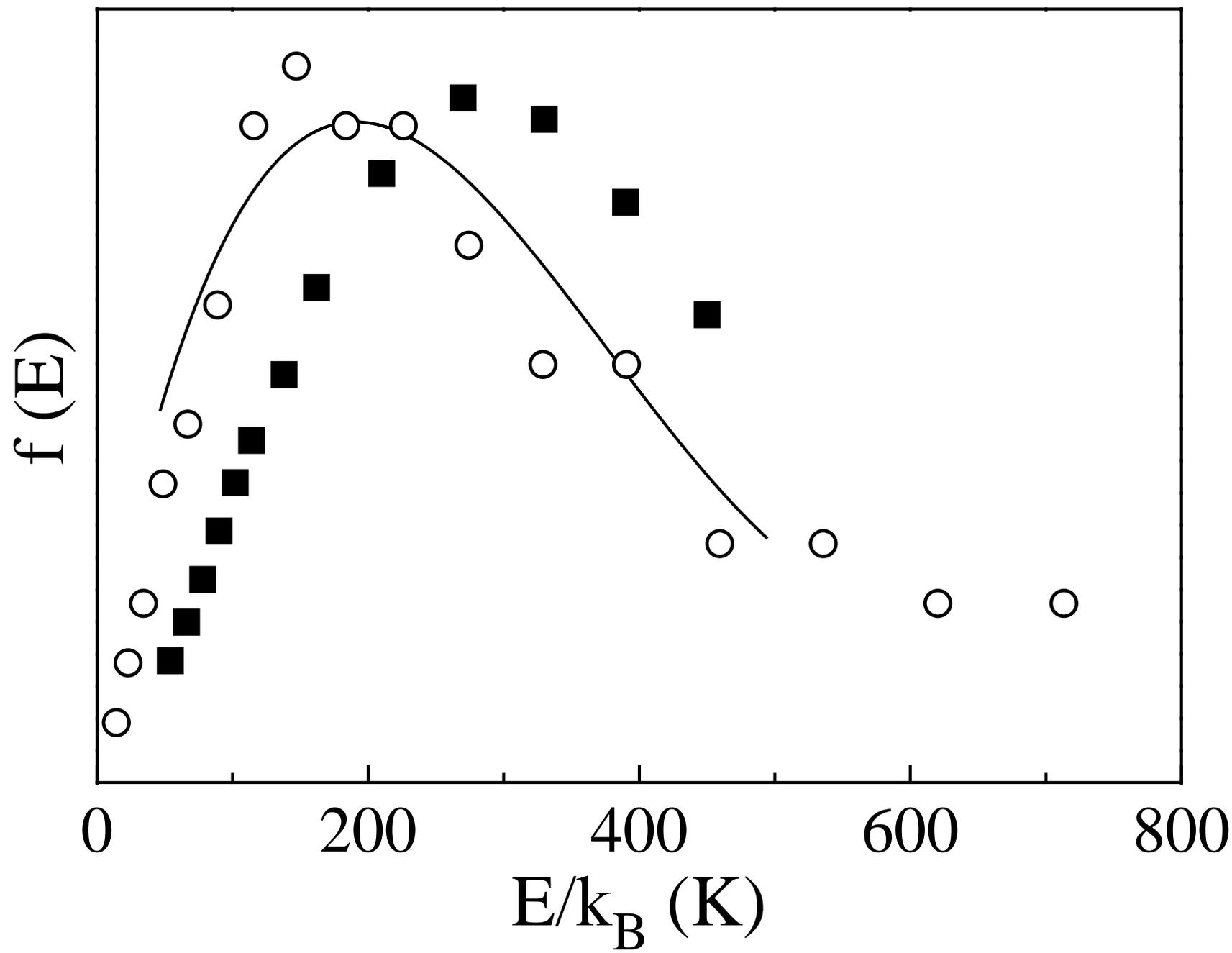
\end{document}